\documentstyle[twoside,fleqn,jhuwkshp,psfig]{article}
\newcommand{\etal}{{\it et al.} }
\newcommand{\asca}{{\it ASCA} }

\newcommand{\xmm}{{\it XMM-Newton} }
\newcommand{\chandra}{{\it Chandra} }

\newcommand{\hetg}{{\it HETGS} }

\newcommand{\fekalfa}{{Fe~K$\alpha$} }

\newcommand{\fexxv}{Fe~{\sc xxv} }
\newcommand{\fexxvp}{Fe~{\sc xxv}}

\newcommand{\feklya}{{Fe~{\sc xxvi }~Ly$\alpha$} }

\newcommand{\feklyap}{{Fe~{\sc xxvi}~Ly$\alpha$}}

\newcommand{\figfinespec}{{Fig.~1} }
\newcommand{\figfinespecp}{{Fig.~1}}
\newcommand{\figivse}{{Fig.~2} }

\newcommand{\figewvsfwhm}{{Fig.~3} }
\newcommand{\figewvsfwhmp}{{Fig.~3} }
\newcommand{\figcoarsesp}{{Fig.~4} }

\newcommand{\tablefits}{Table~1 }
\newcommand{\tablefitsp}{Table~1}
\newcommand{\tablehegfits}{Table~1 }

\begin{document}

\title{The Cores of the Fe K Lines in Seyfert~I
Galaxies Observed by the Chandra High Energy Grating}

\author{Tahir Yaqoob\address{\it Department of Physics and Astronomy,
Johns Hopkins University, Baltimore, MD 21218.}
\address{\it Laboratory for High Energy Astrophysics,
NASA/Goddard Space Flight Center, Greenbelt, MD 20771.} \&
Urmila Padmanabhan$^{\rm a}$
}

\begin{abstract}
\vspace{-6mm}
\centerline{\bf Abstract}

We report on the results of eighteen observations of the
core, or peak, of the \fekalfa emission line at $\sim 6.4$~keV in fifteen
Seyfert~I galaxies using the \chandra High Energy Grating (HEG).
These data afford the highest precision measurements
of the peak energy of the \fekalfa line, and the
highest spectral resolution measurements of the
width of the core of the line
to date. We were able to measure the peak energy in seventeen
data sets, and, excluding a
very deep observation of NGC~3783,
we obtained a weighted mean of $6.404 \pm 0.005$~keV.
In all fifteen sources the two-parameter, 99\% confidence 
errors on the line peak energy do not exclude fluorescent
$K\alpha$
line emission from Fe~{\sc i}, although
two sources (Mkn~509 and 3C~120) stand out as 
very likely being dominated by $K\alpha$ 
emission from Fe~{\sc xvii} or so.
We were able to measure the line core width in fourteen
data sets and obtained a weighted mean of $2380 \pm 760 \ \rm
km \ s^{-1}$ FWHM (excluding the NGC~3783
deep exposure), a little larger than the instrument resolution ($\sim 1860
\ \rm km \ s^{-1}$ FWHM).
However, there is evidence of underlying broad line emission in 
at least four sources. In fact, the width of the peak varies
widely from source to source and it may in general have a
contribution from the outer parts of an accretion disk {\it and}
more distant matter. For the disk contribution to also peak at
6.4~keV requires greater line emissivity at hundreds of gravitational
radii than has been deduced from previous studies of the \fekalfa line.


\begin{center}
{\it Accepted for Publication in the Astrophysical Journal 21 November 2003}
\end{center}

\end{abstract}

\maketitle

\pagebreak

\section{INTRODUCTION}
\label{intro}

\begin{table*}[h]
\caption{\chandra Parameters of the Core Fe K Line Emission from $\it Chandra$ (HEG) Data}
\begin{tabular*}{\textwidth}{@{}l@{\extracolsep{\fill}}cccccc}
\hline
%& & & & & &  \\
Source, & $z^{a}$ & $E^{b}$ & $I^{c}$ & $  {\rm EW}^{d}$ & $ {\rm FWHM}^{e}$
& $ F^{f}$  \\
\& Refs. $^{g}$ & & (keV) & & (eV) & ($\rm km \ s^{-1}$) & $ L^{f}$ \\
%& & & & & & \\
\hline
& & & & & & \\

NGC 7314 & 0.004760 & $6.412^{+0.010}_{-0.015}$ & $1.7^{+0.6}_{-0.8}$ & $48^{+17}_{-23}$ & $100$ f & 3.0 \\
\vspace{-1mm}
$[1]$ & & ($6.395 - 6.430$) & ($0.7 - 2.7$) & ($20 - 76$) & \ldots & $0.15$ \\
NGC 3516(1) & $0.008836$ & $6.398^{+0.017}_{-0.007}$ & $4.0^{+1.3}_{-1.2}$ & $140^{+46}_{-42}$ & $1290^{+1620}_{-1290}$ & $2.3$ \\
\vspace{-1mm}
$[2]$ &    & ($6.389 - 6.408$) & ($2.5 - 5.8$) & ($88 - 203$) & ($0 - 3630$) & $0.40$   \\
NGC 3516(2) & $0.008836$ & $6.401^{+0.015}_{-0.017}$ & $3.8^{+1.5}_{-1.3}$ & $143^{+56}_{-49}$ & $3630^{+2350}_{-1540}$ & $2.1$  \\
$[2]$  &    & ($6.378 - 6.422$) & ($2.2 - 5.8$) & ($83 - 218$) & ($1560 - 7640$) & $0.35$ \\
Mkn 509 & $0.034397$ & $6.430^{+0.024}_{-0.023}$ & $3.1^{+1.9}_{-1.6}$ & $54^{+33}_{-28}$ & $2820^{+2680}_{-2800}$ & $5.3$ \\
\vspace{-1mm}
\ldots   &   & ($6.394 - 6.461$) & ($0.9 - 5.7$) & ($16 - 99$) & ($0 - 7710$) & $14.0$ \\
NGC 5548(1) & $0.016760$ & $6.397^{+0.019}_{-0.023}$ & $3.1^{+1.5}_{-1.2}$ & $115^{+56}_{-45}$ & $3750^{+2590}_{-1890}$ & $2.4$  \\
\vspace{-1mm}
$[3]$  &    & ($6.365 - 6.422$) & ($1.5 - 5.1$) & ($56 - 189$) & ($1250 - 7610$) & $1.5$ \\
NGC 5548(2) & $0.016760$ & $6.400^{+0.010}_{-0.010}$ & $2.1^{+0.8}_{-0.7}$ & $67^{+26}_{-22}$ & $1780^{+1420}_{-1220}$ & $2.9$  \\
\vspace{-1mm}
$[4]$   &    & ($6.386 - 6.414$) & ($1.2 - 3.3$) & ($38 - 105$) & ($0 - 3970$) & $1.8$ \\
3C 120 & $0.033010$ & $6.415^{+0.017}_{-0.017}$ & $3.0^{+1.8}_{-1.5}$ & $61^{+37}_{-31}$ & $2000^{+2950}_{-2000}$ & $4.4$ \\
\vspace{-1mm}
\ldots    &   & ($6.392 - 6.450$) & ($1.1 - 5.5$) & ($22 - 112$) & ($0 - 6770$) & $10.8$ \\
NGC 4593 & $0.008301$ & $6.403^{+0.012}_{-0.038}$ & $3.4^{+3.2}_{-1.3}$ & $80^{+75}_{-31}$ & $2140^{+8370}_{-1230}$ & $4.1$ \\
\vspace{-1mm}
\ldots  &    & ($6.349 - 6.434$) & ($1.8 - 8.1$) & ($42 - 191$) & ($310 - 15920$) & $0.63$ \\
NGC 3783(1) & $0.009730$ & $6.400^{+0.014}_{-0.016}$  & $5.0^{+2.4}_{-2.1}$ & $73^{+35}_{-31}$ & $2550^{+2420}_{-1560}$ & $6.0$ \\
\vspace{-1mm}
$[5]$   &   & ($6.376 - 6.420$) & ($2.4 - 8.3$) & ($35 - 121$) &  ($500 - 6290$) & $1.2$ \\
NGC 3783(2) & $0.009730$ & $6.397^{+0.003}_{-0.003}$ & $4.9^{+0.6}_{-0.5}$ & $70^{+9}_{-7}$ & $1700^{+410}_{-390}$ & $6.2$ \\
\vspace{-1mm}
$[6,7]$    &   & ($6.393 - 6.401$) & ($4.2 - 5.6$) & ($60 - 80$) & ($1180 - 2250$) & $1.3$ \\
MCG -6-30-15 & $0.007749$ & $6.408^{+0.030}_{-0.025}$ & $1.6^{+1.1}_{-0.9}$ & $49^{+34}_{-28}$ & $3250^{+5230}_{-3250}$ & $3.2$  \\
\vspace{-1mm}
$[8]$  &    & ($6.349 - 6.454$) & ($0.4 - 3.3$) & ($12 - 101$) & ($0 - 12670$) & $0.42$ \\
Mkn 279 & $0.030451$ & $6.415^{+0.047}_{-0.027}$  & $1.9^{+1.0}_{-0.9}$ & $132^{+69}_{-63}$ & $5010^{+6550}_{-2810}$ & $1.3$  \\
\vspace{-1mm}
\ldots  &   & ($6.379-6.521$) & ($0.7-5.2$) & ($49-361$) &  ($1100-36980$) & $2.7$ \\
NGC 4051 & $0.002336$ & $6.419^{+0.038}_{-0.033}$ & $3.2^{+1.7}_{-1.4}$ & $191^{+101}_{-84}$ & $6330^{+7740}_{-3330}$ & $1.6$ \\
\vspace{-1mm}
$[9]$   &   & ($ 5.910 - 6.475$) & ($1.4 - 25.0$) & ($84 - 1492$) & ($1780 - 137470$) & $0.019$ \\
IC 4329A & $0.016094$ & $6.309^{+0.089}_{-0.099}$ & $11.3^{+7.2}_{-6.4}$ & $62^{+40}_{-35}$ & $15090^{+12430}_{-9950}$ & $16.4$  \\
\vspace{-1mm}
 \ldots  &    & ($6.171 -6.468$) & ($1.8 - 21.3$) & ($10 - 117$) & ($0 - 37150$) & $9.4$ \\
F 9 & $0.047016$ & $6.373^{+0.254}_{-0.092}$ & $4.9^{+8.7}_{-3.3}$ & $216^{+384}_{-145}$ & $17040^{+55960}_{-14270}$ & $2.1$  \\
\vspace{-1mm}
\ldots  &  & ($6.203 - 6.753$) & ($1.2 - 17.0$) & ($53 - 749$) & ($2160 - 92940$) & $10.6$ \\
Mkn 766 & $0.012929$ & $6.423^{+0.018}_{-0.016}$ & $0.7^{+0.6}_{-0.5}$ & $34^{+29}_{-24}$ 
& $100$ f & $2.1$  \\
\vspace{-1mm}
\ldots   &    & ($6.398 - 6.456$) & ($0.1 - 1.5$) & ($5 - 73$) & \ldots & $0.79$ \\
NGC 3227 & $0.003859$ & $6.384^{+0.015}_{-0.016}$ & $1.1^{+1.2}_{-0.9}$ & $39^{+43}_{-32}$ 
& $100$ f & $2.3$  \\
\vspace{-1mm}
\ldots   &    & \ldots & ($0 - 2.8$) & ($0 - 99$) & \ldots & $0.075$ \\
Akn 564 & $0.024684$ & $6.400$ f & $0.3^{+0.5}_{-0.3}$ & $16^{+27}_{-16}$ & $100$f & $2.5$  \\
\vspace{-1mm}
$[10]$   &    & \ldots & ($0.0 - 1.1$) & ($0 - 59$) & \ldots & $3.4$ \\
& & & & & & \\
\hline
\end{tabular*}
{\small
{\it Chandra} HEG data, fitted with a power law plus Gaussian emission line model 
in the 2--7 keV band
(for NGC 3516 photoelectric absorption was also included in the model).
All parameters (except redshift) were free in the fits, except in the cases of 
NGC~7314, Mkn~766, NGC 3227 (line width fixed in these cases), and Akn 564 (line energy
and width fixed). All parameters are quoted in the source rest
frame. 
Statistical errors are for the 68\% confidence
level, whilst parentheses show the
90\% confidence level ranges of the parameters.
The number of interesting parameters assumed for calculating the
statistical errors was equal to the number of free parameters
in the Gaussian component of the model.
For 3, 2, and 1 interesting parameter(s), the corresponding 
values of $\Delta C$ for 68\% confidence are  3.506, 
2.279 and 0.989 respectively. For 3, 2, and 1 interesting parameter(s), the 90\% 
confidence values of
$\Delta C$  are 6.251, 4.605 and 2.706 respectively. 
$^{a}$  Redshifts obtained from NASA Extragalactic Database (NED).
$^{b}$  Gaussian line center energy.   
$^{c}$  Emission-line intensity in units of $\rm 10^{-5} \ photons \ cm^{-2} \ s^{-1}$.
$^{d}$  Emission line equivalent width.
$^{e}$  Full width half maximum, rounded to $10 \ \rm km \ s^{-1}$.
$^{f}$  $F$ is the 
estimated 2--10~keV observed flux in units of $10^{-11} \ \rm ergs\ cm^{-2}\ s^{-1}$.
The power-law continuum was extrapolated to 10 keV.
$L$ is the estimated 2--10~keV source-frame luminosity
(using the 2--10 keV estimated flux), in units of $10^{43} \ \rm ergs\ s^{-1}$, assuming
$H_{0} = \rm 70 \ km \ s^{-1} \ Mpc^{-1}$ and $q_{0}$.
$^{g}$ Key for previous publications on the same HEG data:
[1] Yaqoob \etal 2003a, [2] Turner \etal 2002, [3] Yaqoob \etal 2001,
[4] Kaastra \etal 2002, [5] Kaspi \etal 2001, [6] Kaspi \etal 2002, [7] Netzer \etal 2003,
[8] Lee \etal 2002, [9] Collinge \etal 2001, [10] Mastumoto, Leighly, \& Marshall 2001,
http://www.pha.jhu.edu/groups/astro/workshop2001/papers/matsumoto\_c.ps}.
\end{table*}

\begin{figure*}[h]
\vspace{10pt}
\centerline{\psfig{file=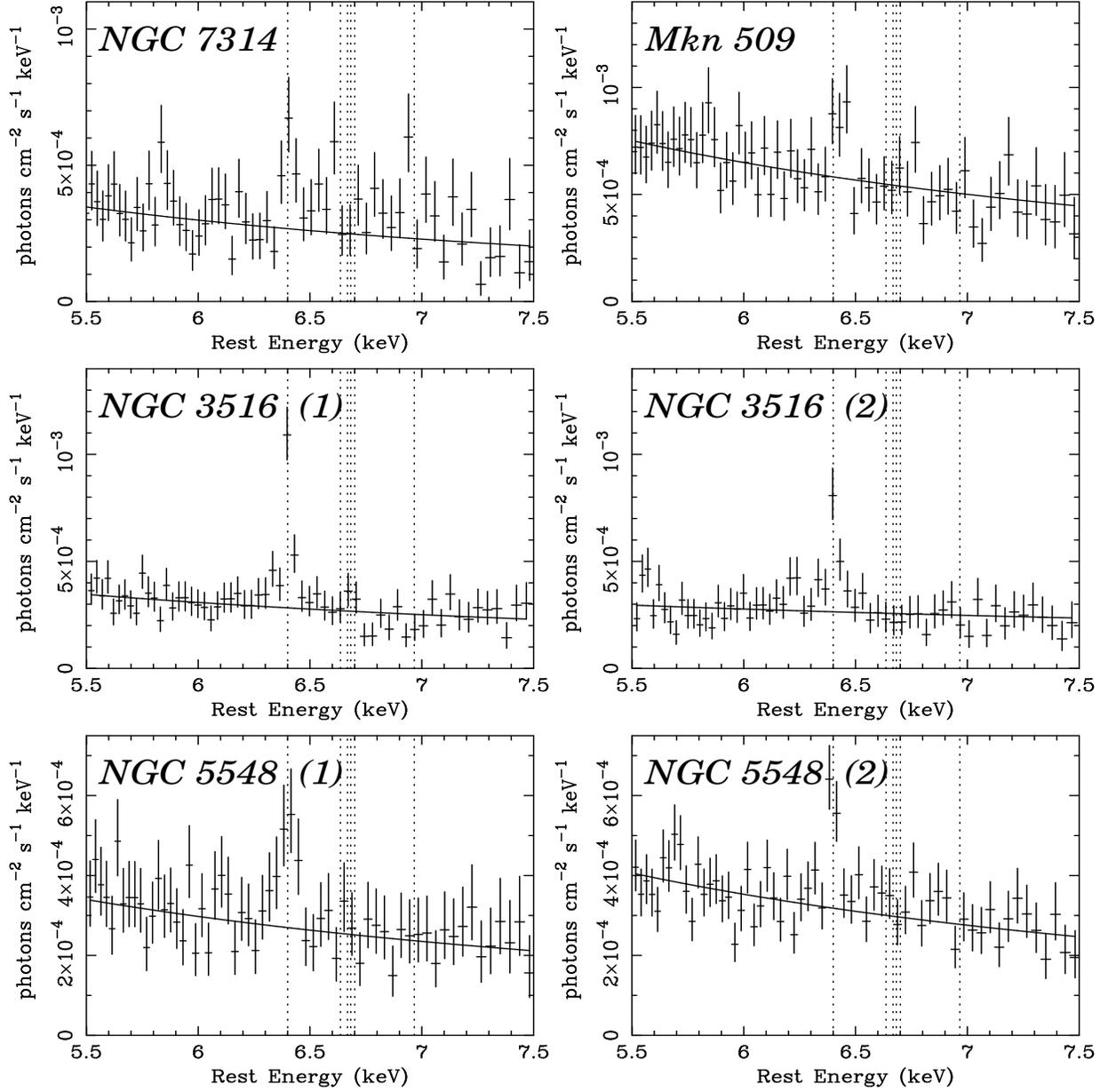,width=6.5in,height=6.5in,angle=180}}
%\centerline{\psfig{file=f1b.eps,width=9.0in,height=8.0in,angle=180}}
%\centerline{\psfig{file=f1c.eps,width=9.0in,height=8.0in,angle=180}}
\caption{\footnotesize 
\chandra HEG spectra in the Fe K band from each of the eighteen
observations of fifteen type~I AGN (see \tablefitsp).
The data are binned at $0.01\AA$ for the first sixteen spectra
shown, and $0.02\AA$ for the last two spectra (NGC~3227 \& Akn~564).
This can be compared to the HEG spectral resolution, which is $0.012\AA$
FWHM. The data are combined from the $-1$ and $+1$ orders
of the grating. The spectra have been corrected for instrumental effective area
and cosmological redshift.
Note that these are {\it not} unfolded spectra and are
therefore independent of the model that is fitted.
The statistical errors shown correspond to the $1\sigma$ 
Poisson errors, calculated using 
equations (7) and (14) in Geherls (1986) to approximate the upper and
lower errors respectively.
The solid line corresponds to the continuum model fitted over the
2--7~keV range (extrapolated to 7.5~keV), as described in the text (\S\ref{hegspec}).
The vertical dotted lines represent (from left to right), the rest energies
of the following transitions: Fe~{\sc i}~$K\alpha$, \fexxv forbidden,
two intercombination lines of \fexxvp, \fexxv resonance, and \feklyap.
The spectrum for NGC~7314 corresponds to that during a low continuum
state, as defined and described in Yaqoob \etal 2003a.
Note that an apparent narrow feature at
$\sim 6.35$~keV in MCG~$-$6-30-15 and at $\sim 6.55$~keV in
NGC~3783(2) are detected in one arm of the grating only and
are narrower than the spectral resolution, so are not real.}
\end{figure*}

\setcounter{figure}{0}
\begin{figure*}[tbh]
\vspace{10pt}
\centerline{\psfig{file=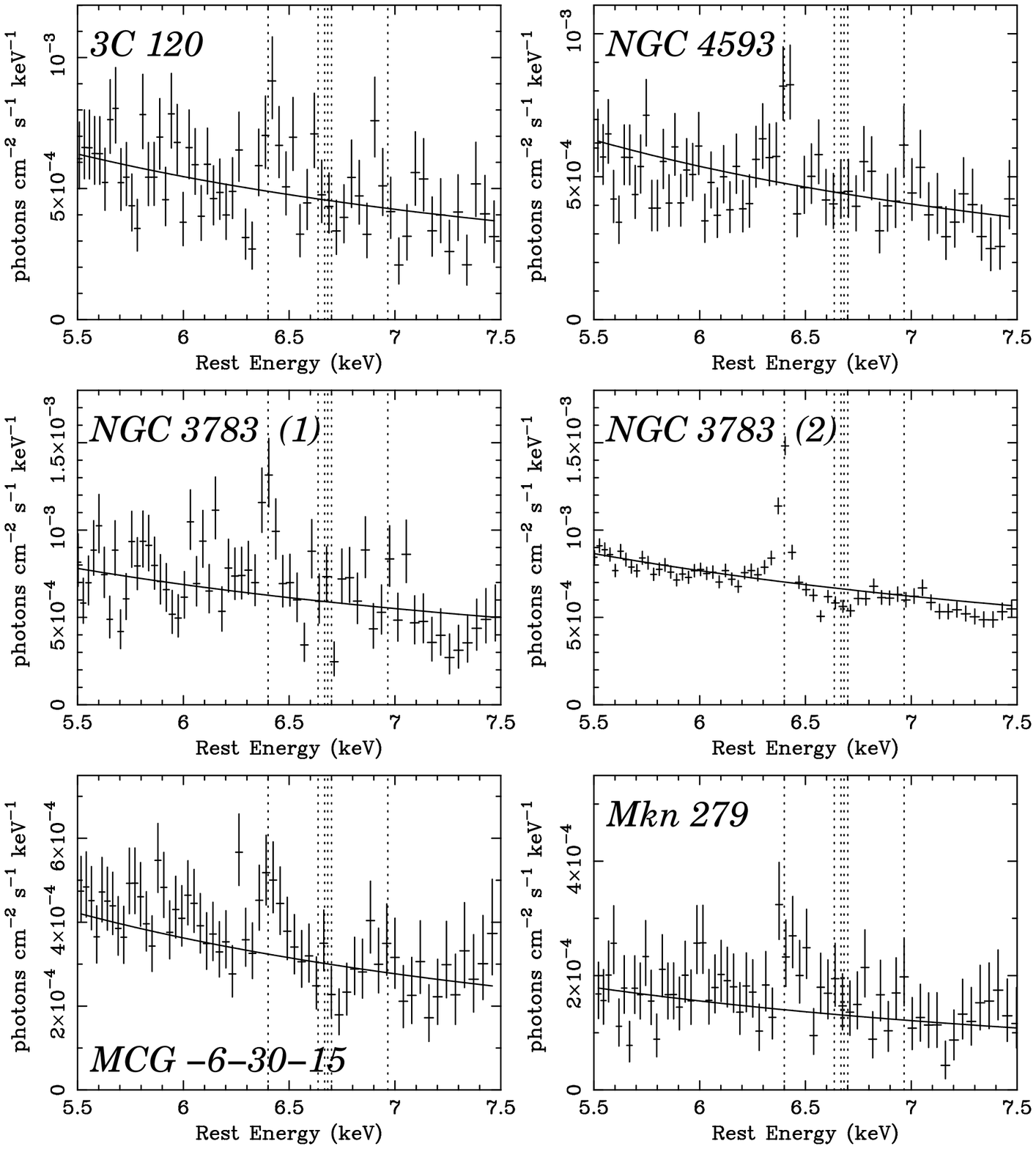,width=6.5in,height=6.5in,angle=180}}
\caption{ -- {\it continued}}
\end{figure*}

\setcounter{figure}{0}
\begin{figure*}[tbh]
\vspace{10pt}
\centerline{\psfig{file=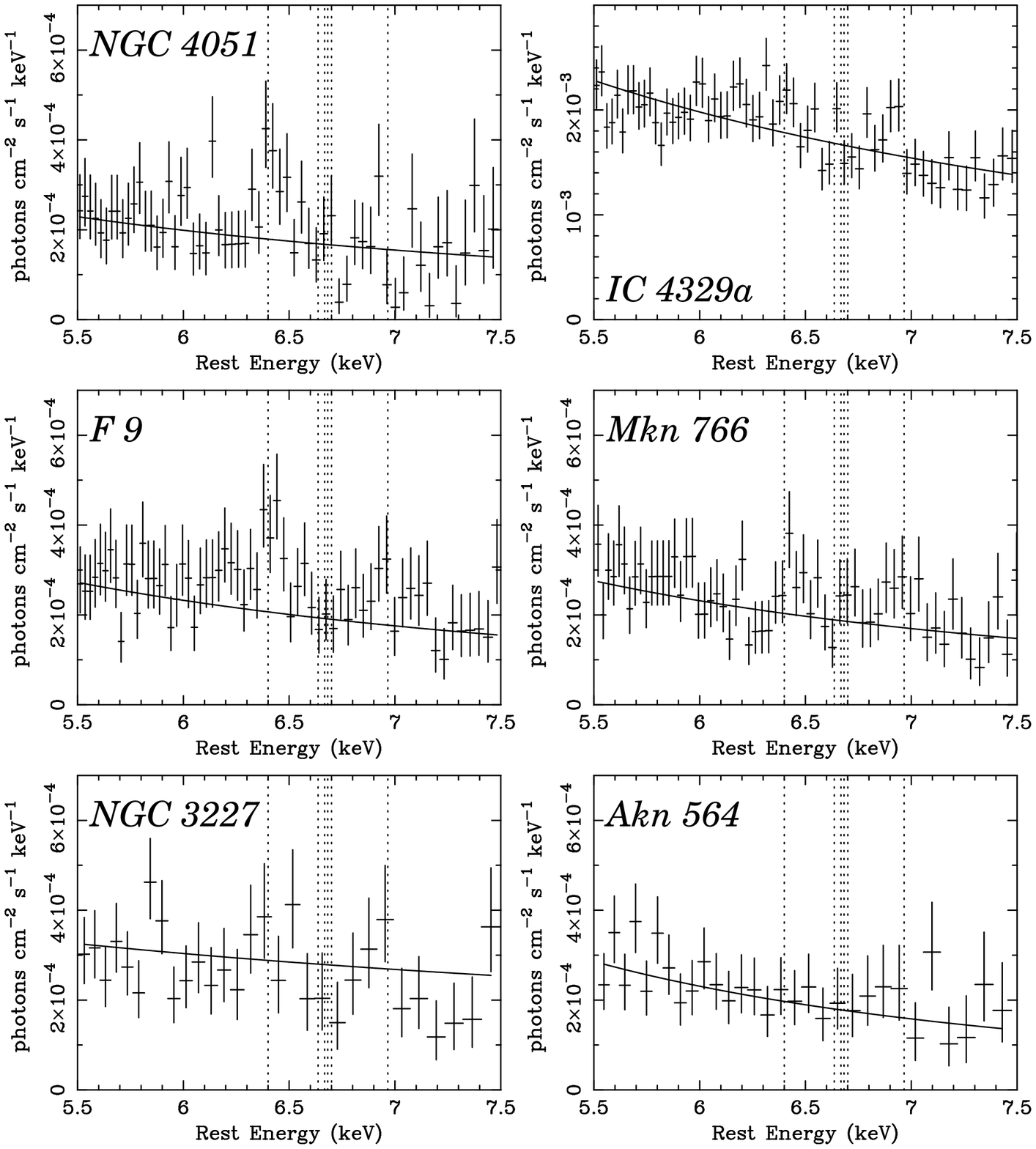,width=6.5in,height=6.5in,angle=180}}
\caption{ -- {\it continued}}
\end{figure*}

At least part of the 
\fekalfa fluorescent 
emission line in type~I active galactic nuclei (AGNs)
is believed to originate in a
relativistic accretion disk around a black hole
(e.g. see reviews by Fabian \etal 2000; Reynolds \& Nowak 2003).
The dominant peak energy of the \fekalfa line 
at $\sim 6.4$~keV 
appears to be ubiquitous and this core of the line carries a
substantial fraction of the total line flux 
(e.g. Nandra \etal 1997a; Sulentic \etal 1998; Lubi\'{n}ski \& Zdziarski 2001; 
Weaver, Gelbord, \& Yaqoob 2001; Yaqoob \etal 2002;
Perola \etal 2002; Reeves 2002). Often, the broad part
of the \fekalfa line is absent, leaving only the narrow core. 
It has been traditional to associate such narrow \fekalfa lines   
with an origin in distant matter, at least several
thousand gravitational radii from the putative black hole (e.g.
the optical broad-line region (BLR), the
putative obscuring torus, or the optical narrow-line region (NLR)).
However, Petrucci \etal (2002) recently reported a {\it variable},
narrow \fekalfa line in Mkn~841, supporting an accretion-disk origin.
Moreover, rapidly variable, narrow Fe K line emission 
has been observed in the Seyfert~I galaxy NGC~7314 (Yaqoob \etal 2003a).
Thus, even narrow \fekalfa lines may have
a significant contribution from the accretion disk
(Lee \etal 2002; Yaqoob \etal 2003a).
While such lines may be interpreted in terms of
a truncated disk (e.g. Done, Madejski, \& \.{Z}ycki 2000), they could be
due to low-inclination angle disks with a flat radial line emissivity (i.e. 
intensity per unit area falling off with radius more
slowly than $r^{-2}$). 

In this paper we address two very specific questions,
using the \chandra high energy grating transmission spectrometer ({\it HETGS};
see Markert \etal 1995), which affords the best spectral
resolution currently available at 6.4~keV 
($\sim 39$~eV, or $\sim 1860 \ \rm km \ s^{-1}$ FWHM).
Namely, for a sample of type~I AGNs,
what are the energies of the peaks of the Fe K line emission and
are these line cores resolved by {\it Chandra}?
This information can give important clues about the ionization state
of Fe responsible for the line emission, and its origin.
With \chandra we can measure the peak energies with better
precision than {\it ASCA}, at least by a factor of four. 

We note that, due to the small throughput of the \hetg 
(especially above $\sim 7$~keV), it is very difficult for the grating
data to constrain the parameters of any underlying broad \fekalfa line
emission. However, in a different study we shall
systematically compare the total \fekalfa line emission
observed with \chandra and \asca data and show that, aside from variability
in some sources, there is good agreement between the two sets of data.
We emphasize that even though the \chandra data for most
type~I AGN show narrow \fekalfa line peaks, this by no means indicates that
there is no broad \fekalfa line emission. 

\section{OBSERVATIONS AND DATA}
\label{data}
 
Our study is based on eighteen observations of fifteen type~I AGN 
(see \tablefitsp) with $z<0.05$
observed with {\it HETGS}, that were in the \chandra public
archives as of 2003, July 1, and had a 
total, first-order, HEG count rate higher than
0.05 ct/s. Blazars, BL~Lac objects, and AGN that are intermediate
between type~I and type~II were excluded from the study.
Details of all the observations can be found from the \chandra public
archive\footnote{http://cda.harvard.edu:9011/chaser/}.
Three sources were observed more than once: NGC~3516, NGC~5548, and
NGC~3783. For NGC~3516, observation IDs 2080 and 2431 were combined
since they occurred on consecutive days, whilst observation ID 2482 was treated
separately (see also Turner \etal 2002). 
The two observations of NGC~5548 were treated separately.
For NGC~3783, six observations were at first treated separately
but we found that the Fe~K line parameters were consistent
with no variability, so here we report the results from the
$\sim 850$~ks spectrum 
combined from five snapshots
taken during a monitoring campaign but treated this separately
from an observation made a year earlier. Detailed results from the
monitoring campaign will be presented elsewhere (Yaqoob {\it et al.},
in preparation; see also Kaspi \etal 2002, and Netzer \etal 2003).

\hetg consists of two grating assemblies,
a High-Energy Grating (HEG) and a Medium-Energy Grating (MEG),
and it is the HEG that achieves the highest spectral resolution.
The MEG has only half of the spectral resolution
of the HEG and less effective area in the Fe-K band, so our study will
focus on the HEG data.

The \chandra data 
were reduced and HEG spectra made, exactly as described in Yaqoob \etal (2003b).
We used only the first orders of the grating data (combining
the positive and negative arms). 
The mean HEG count rates ranged from $0.087$ ct/s
for the weakest source (NGC~3227) to $1.16 \pm 0.006$ for
the brightest source (IC~4329a).
The exposure time ranged from $\sim 40$~ks to $\sim 850$~ks, but
was $\sim 60-100$~ks for most of the sources.
Background was not subtracted since it is negligible over the
energy range of interest (e.g. see Yaqoob \etal 2003a).
Note that 
the systematic uncertainty in the HEG
wavelength scale 
is $\sim 433 \ \rm km \ s^{-1}$ ($\sim 11$~eV) at 6.4 keV
\footnote{http://space.mit.edu/CXC/calib/hetgcal.html}.

\section{SPECTRAL FITTING RESULTS}
\label{hegspec}

We used XSPEC v11.2 (Arnaud 1996) for spectral fitting.
Since we are interested in utilizing the highest possible
spectral resolution available, we used spectra binned
at $0.0025\AA$, and this amply oversamples the HEG resolution
($0.012\AA$ FWHM).
The $C$-statistic was used for minimization.
All model parameters will be
referred to the source frame.
Our method is simply to fit a simple power-law plus Gaussian
emission-line model over the 2--7~keV band for each spectrum.
NGC~3516 required photoelectric absorption to fit the continuum.
Also, for NGC~7314 the analysis was more complex,
involving emission from multiple ionization states of Fe, and has already
been described in detail in Yaqoob \etal (2003a). The results
presented here for NGC~7314 are for the 
6.4~keV line component only (which is unresolved) and have been taken from 
Yaqoob \etal (2003a), changing only
the confidence levels of the quoted statistical errors so
that they are consistent with the rest of the sample.
For Akn~564 an
Fe K emission line was not detected:
$C$ decreased by 0.9 ($<68\%$ confidence for the
addition of one free parameter) when a
narrow (FWHM much less than the instrument resolution) emission line at 6.4~keV was
added to a power-law only. Thus, we 
obtained upper limits on the equivalent width (EW).
A significant Fe K emission line has been detected in Akn~564 with
\asca (Turner \etal 2001). The reason for the non-detection by the HEG
is likely to be due to the very steep continuum and the small effective
area of the HEG. The line was only weakly 
detected during recent \xmm observations (Vignali \etal 2003),
but the signal-to-noise
of those data was still much less than that of the \asca data. 
In the case of NGC~3227 and Mkn~766 the detection of an Fe K line was marginal:
$C$ decreased by 4.5 and 7.0 respectively, when a narrow Gaussian 
was added to a power-law model only. In this case we were able
to obtain constraints on the line energy and EW, so the 
Gaussian model had two free parameters. Thus, the lines were
detected with $<90\%$ and $<95\%$ confidence in NGC~3227 and
Mkn~766 respectively. Note that
a strong Fe K line has been detected in NGC~3227 by \xmm (Gondoin \etal 2003),
and complex Fe K emission
has been observed in Mkn~766 by \xmm (Pounds \etal 2003).
For the remaining spectra the Gaussian component had three free
parameters (line center energy, width, and intensity).
Thus, except for the three cases mentioned above, the model had
five free parameters in total, including the continuum slope
and normalization. 

We used the `goodness' command in XSPEC to assess
the goodness of the fits: this command performs
Monte Carlo simulations of
spectra using the best-fitting model and gives the
percentage of the simulated spectra that had a
fit statistic less than that obtained from the fit
to the real data. A value of 50\% is expected if
the best-fitting model is a good representation of the
data. Values much less than 50\% indicate that the
data are over-parameterized by the model since
random statistical fluctuations in the majority of the simulated
spectra are not able to produce a fit statistic as
low as that obtained from the real data. In the
opposite limit,
when 100\% of the simulated spectra have a fit
statistic less than that obtained from the real data,
the fit is clearly poor.  

Good fits were obtained for all spectra except for
NGC~3783(2), the long $\sim 850$~ks observation, in which the  
continuum is complicated by a warm absorber that affects
the spectrum even above 2~keV (e.g. see Kaspi \etal 2002, and
Netzer \etal 2003). However, this has little impact on the
deduced parameters of the \fekalfa line which are in fact
consistent with those obtained by Kaspi \etal (2002) and
Netzer \etal (2003), who used a more complex continuum.
The line parameters, including the intensity, are also
consistent with those measured from a non-simultaneous \xmm
observation (Reeves \etal 2004), for which a warm absorber
was also included in the continuum modeling.

Excluding Akn~564, NGC~3227, and Mkn~766, the \fekalfa
line core was detected at a confidence level
$>3\sigma$ (corresponding to a decrease in $C$ greater
than 14.16 when a Gaussian with three free parameters 
was added to the continuum).
Five of these fifteen spectra gave `goodness' values less than 50\%.
Two of these were close to 50\% (NGC~4051: 49\%, F~9: 47\%)
so the data are likely not over-parameterized. Also,
$C$ decreased by 33.2 and 26.6 for NGC~4051 and F~9 respectively,
when a three-parameter Gaussian model was added to the
continuum model. Thus the line-emission was detected 
at a confidence level $>5\sigma$ and $>4\sigma$ in
NGC~4051 and F~9 respectively.
In the other three cases, the goodness values were
32\% (Mkn~509), 10\% (NGC~3516(2)), and 7\%(NGC~5548(1)).
In the case of Mkn~509, when the data were modeled without
an emission line, the `goodness' value was 40\%, still
less than 50\%,
indicating that the signal-to-noise over the entire energy band
is poor. The decrease in $C$ when a Gaussian is added to
the continuum is 16.8, which still corresponds to a 
detection at a confidence level $>3\sigma$, for the
addition of three free parameters. For NGC~3516(1),
the `goodness' value increases from 10\% to 69\% when the
emission line is removed and the data fitted with a
continuum only. The addition of a three-parameter Gaussian
to the continuum decreases $C$ by 111.4, confirming what
is evident from the spectrum in \figfinespecp, that the
line emission is highly significant and required by the data.
In the case of NGC~5548(1), removing the emission line
and fitting a continuum only, increases the `goodness' value
from 7\% to 18\%, still below 50\%. However, the addition of
a three-parameter Gaussian to the continuum decreases $C$
by 35.0, indicating that the \fekalfa line is detected at
a confidence level $>5\sigma$. The low values of the
`goodness' parameter are due to poor signal-to-noise over
the whole energy band since a simple two-parameter continuum
gives a `goodness' value much less than 50\%.

Although the \fekalfa line consists of two components
($K\alpha_{1}$ and $K\alpha_{2}$, separated by 13~eV), we
modeled it as a single Gaussian, since it was shown in Yaqoob \etal (2001),
that with the spectral resolution of the HEG, there
is a negligible impact on the measured line width. 
Some broadening may also result from the presence of
line emission from more than one ionization state of Fe. However, we
do not interpret the measured FWHM velocities literally.
Also, the use of a single Gaussian (without any attempt to
model the underlying broad \fekalfa emission) also has a negligible
impact on the measured center energy of the core, the line intensity
and EW (see Yaqoob \etal 2001). Again, we only
interpret the line intensity and EW qualitatively.
Furthermore, one of the reasons for
measuring the width of the line core is to obtain clues about any 
underlying broad \fekalfa line component.

The best-fitting emission-line parameters for each spectrum
are shown in \tablefits (as well as extrapolated 2--10~keV
fluxes and luminosities). Note that since the models were fitted
by first folding through the instrument response before comparing
with the data, the derived line parameters {\it do not}
need to be corrected for instrumental response.
In order that the results can used for future statistical analyses
the statistical errors shown 
correspond to 68\% confidence 
($\Delta C = 3.506, 2.279$, or 0.989, depending on whether
there were 3, 2 or 1 free parameter(s) free in
the Gaussian component). However, as a more conservative measure,
the 90\% confidence range for each line parameter is also given 
in \tablefitsp.
\figfinespec shows each of the eighteen spectra in the Fe K region,
corrected for instrumental efficiency and cosomological redshift.
The spectra are binned at $0.01\AA$, similar to the HEG spectral
resolution of $0.012\AA$, so broad features are not readily
discernable in this representation. 

\section{PROPERTIES OF THE CORE OF THE FE K LINE EMISSION}
\label{properties}

\figivse shows joint, 99\% confidence contours of the line
intensity versus line center energy for thirteen of the eighteen spectra,
and \figewvsfwhm shows the 99\% confidence contours
of the line EW versus FWHM width for the same spectra.
Excluded were Mkn~766, NGC~3227, and Akn~564, 
(since the constraints on the line parameters are poor),
the short, first
observation of NGC~3783 (since the later observation had $\sim 20$
times the exposure time), and NGC~7314 (that has complex
Fe~K emission from multiple ionization states: see Yaqoob \etal 2003a).

\figivse and \figewvsfwhm show the results for the eleven sources,
split into two groups. 
Group~1 (\figivse(a) and \figewvsfwhm(a))
is comprised of NGC~3516, Mkn~509, NGC~5548, 3C~120,
NGC~3783, and MCG~$-$6-30-15.
Group~2 (\figivse(b) and \figewvsfwhm(b))
is comprised of NGC~4593, Mkn~279, NGC~4051, IC~4329a, and F~9.
Roughly speaking, group~1 AGN have more prominent narrow Fe K line
cores than group~2, as evidenced by the larger contours for
group~2. In fact, for NGC~4593 and NGC~4051 the 99\% contours
could not be well constrained because when the
Gaussian is very broad, there is a lot of interaction between
the line width and the continuum slope,
particularly if there is a reflection continuum, which we
have not modeled here. Therefore we constructed
additional contours for all
the sources by fixing the power-law slope at the best-fitting value
for each source. These contours are shown with dashed lines in
\figivse(b) and \figewvsfwhm(b). For the group~1 sources, the
differences between the two sets of contours were negligible so they
are not shown for the sake of clarity.

We emphasize that the size of the 99\% contours is not simply
a function of signal-to-noise of the data. To illustrate this
point, we have included a group~1 contour (NGC~5548, first observation)
in the group 2 plots. We can compare this with F~9, that has the
largest contours of all eleven sources. Now, the total number of photons in the 5--7~keV
band
in the NGC~5548 and F~9 spectra is 1240 and 1146 respectively.
Thus, the significant differences in the sizes of the contours
for the two sources are due to intrinsic differences in the emission-line
profile, not just signal-to-noise. Furthermore, since NGC~5548 and F~9
have the lowest signal-to-noise spectra of the eleven sources, we can
conclude that the relative differences in the contours
in general may be 
due to intrinsic differences in the line shape.

Physically, what this means is that the group~1 sources
have Fe K line emission, that at the {\it very peak} is dominated
by a low-velocity emission component
that is near the rest-energy of Fe~{\sc i}~$K\alpha$. This does
not, of course, mean that group~1 sources do not necessarily have 
broad \fekalfa line components. Indeed, MCG~$-$6-30-15 is in group~1, but
it has the strongest and broadest \fekalfa line yet observed in an AGN.
It simply means that our single-Gaussian fits are picking up
a narrow component at $\sim 6.4$~keV that stands prominently
above the underlying broad line because the latter is spread
out over such a large energy range. In the group~2 sources,
the {\it peak} line emission at $\sim 6.4$~keV is not dominated
by a narrow component, but has a significant contribution from
an underlying broad component. This could be because the narrow
component is weaker relative to the
broad component than it is in group~1, {\it or} it could mean
that the broad component is narrower than 
it is in group~1.

We obtained a weighted mean 
line center energy of $6.399 \pm 0.003$~keV for the seventeen out of
eighteen spectra in which the line energy could be measured
(see \tablefitsp). Omitting 
NGC~3783(2) to avoid bias due to the very deep, $\sim 850$~ks exposure of this AGN,
we obtained $6.404 \pm 0.005$~keV. 
Here, for the calculation of the weighted mean of
any quantity with asymmetric errors, we
simply assume symmetric errors, using the largest 68\% confidence
error in \tablefitsp.
The weighted mean FWHM of the \fekalfa line cores for the fourteen 
data sets for which it could be measured, is $1850 \pm 360 \ \rm 
km \ s^{-1}$. Without NGC~3783(2) it is $2380 \pm 760 \ \rm
km \ s^{-1}$.
For the eight group~1 observations of
six AGN (see \figivse(a) and \figewvsfwhm(a)),
the weighted mean line center energy and
FWHM is $6.398 \pm 0.003$~keV and $1756 \pm 366 \rm \ km \ s^{-1}$
respectively.
For the five observations of
the five sources in group~2 (see \figivse(b) and \figewvsfwhm(b)),
the weighted mean line center energy and
FWHM is $6.406 \pm 0.023$ and $5831 \pm 4046 \rm \ km \ s^{-1}$
respectively.
At 99\% confidence, the \chandra HEG resolves the narrow
component of the \fekalfa emission in three group~1 spectra
(NGC~3516(2), NGC~3783(2), and NGC~5548(1)), and
three group~2 spectra (Mkn~279, NGC~4051, and F~9).
Interestingly, in  NGC~3516
and NGC~5548 (which have multiple observations), 
the line is resolved in the {\it lower} signal-to-noise
spectrum in each case. In each source the continuum level
appears to be similar for the pair of observations. 
Therefore there appears to be real variability in the
line width on a timescale of months to years, which indicates
a change in the dominant distance of the line emission relative to
the putative central black hole. Alternatively,
it may be that there is a variable broad accretion-disk component 
affecting the measured width of the line core.  

\figivse(a) shows that for NGC~3516, NGC~3783, and NGC~5548
the 99\% contours of line intensity versus center energy
are less than $\sim 80$~eV wide
and are fairly symmetrical about 6.400~keV. Thus, in these AGN,
the narrow \fekalfa component detected by the \hetg
is predominantly from distant matter. The FWHM contours in
\figewvsfwhm suggest an origin at the location of the
optical BLR and/or beyond (see also Yaqoob \etal 2001; Kaspi \etal 2002;
Page \etal 2003). Due to the symmetry of the contours
around 6.400~keV, the 99\% confidence bounds imply
that most of the Fe must be less ionized than Fe~{\sc xv} or
so, with Fe~{\sc i} being the most likely ionization state.
For 3C~120 and Mkn~509, the 99\% intensity versus energy
contours are $\sim 70$ and $\sim 90$ eV wide, respectively,
and both contours are centered significantly above 6.400 keV
(by $\sim 25$~eV). For these two sources, the most 
dominant ionization stage is likely to be Fe~{\sc xvii} or so, but the
99\% contours do not rule out anything in the range  Fe~{\sc i}
to  Fe~{\sc xix}. We note that 3C~120 and Mkn~509 are
two of the most luminous sources in the sample
(see \tablefitsp) so a higher ionization state for the line-emitting
matter may be commensurate with this (e.g. see Nandra \etal 1997b).
Thus, not only is the origin of the entire \fekalfa line emission in these
two sources controversial (3C 120: see Zdziarski \& Grandi 2001;
Mkn~509: see Pounds \etal 2001; Page, Davis, \& Salvi 2003; De Rosa \etal 2003),
the origin of the peak emission is also ambiguous.

The last source, MCG$-$6-30-15, has the widest
intensity versus energy contour ($\sim 170$~eV) in group~1,
so the narrow \fekalfa line component in this case clearly
is being affected by the underlying broad line. 
Since the contour is symmetric about 6.4~keV, the ionization state
is likely to be low (but the 99\% upper limit allow 
ionization states up to Fe~{\sc xvii}). We note that the EW of the
narrow-line component is about the same as that obtained by
Wilms \etal (2001) and Ballantyne, Ross, \& Fabian (2001) who included it
in a complex broad plus narrow-line model applied to \xmm data. 

Strictly speaking, in all six of the group~1 sources, we cannot
rule out higher ionization states than mentioned,
due to the possibility of 
gravitational redshifts affecting the line peak. However, it
seems unlikely that the dominant ionization stage of Fe and the
gravitational redshifting would conspire to give a center energy
so close to 6.4~keV in four of the six group~1 sources.
 
Of the five group~2 AGN, the \fekalfa line cores in three of them
are resolved (Mkn~279, NGC~4051, and F~9). This, along with the fact that
the 99\% contours in both center energy and line width (\figivse(b) and
\figewvsfwhm(b) respectively) are very wide, implies that 
the line cores in these
sources are clearly dominated by a broad line. However, the fact that
the line center best-fitting values are still close to 6.4~keV,
and the presence of asymmetry in the contours (compared to group~1),
means that there is still an important contribution from
matter that is in the outer accretion disk or far from it.
In NGC~4593 and IC~4329a the line {\it peak} is not resolved and
is stronger than in the other three group~2 sources, but not
as dominant over the broad \fekalfa emission as it is for the group~1
sources.

\figcoarsesp shows the ratios of the spectral data to
the simple fitted continuua, in the 3--8~keV band for
the group~2 sources. Note that MCG~$-$6-30-15 is
a group~1 source but still has a strong broad Fe K line
component. However, it is not shown in \figcoarsesp
since a similar plot has 
already been shown by Lee \etal (2002) and discussed at length.
The data in \figcoarsesp have been binned at $0.04\AA$ so the
broad structure of the Fe K line emission in these sources is  
more readily apparent than in \figfinespecp.
However, we find that, except for IC~4329a, the signal-to-noise
and bandpass of the data are insufficient to provide useful
constraints on physical models from spectral fitting.
In any case, this is beyond the scope of the present paper, in 
which we are concerned with measurements of the line cores.
IC~4329a is by far the brightest AGN in the sample and
detailed modeling of the complex Fe K line emission apparent
in \figcoarsesp is presented in McKernan, Yaqoob, \& Padmanabhan (2004,
in preparation). Here we simply note that one interpretation of the
data for IC~4329a is that the higher-energy peak in the Fe K complex
is due to \feklya emission. NGC~7314 is the only other source in this
\chandra sample with a statistically significant 
peak near the energy expected for \feklya (see Yaqoob \etal 2003a). 
\xmm data for MCG~$-$6-30-15 show structure in the line
emission above 6.4~keV that could either be interpreted
as Fe He-like resonance absorption at $\sim 6.7$~keV, or
Fe H-like emission at $\sim 6.9$~keV
(Fabian \etal 2002). The HEG data are rather ambiguous.
He-like absorption may be present but the
signal-to-noise is too poor
(see also Lee \etal 2002).
In our HEG sample, there is marginal evidence for a peak near
$\sim 6.9$~keV in 3C~120, NGC~4593, F~9, and Mkn~766.
Aside from NGC~7314, emission from He-like Fe is not evident from
any of the other
spectra. It is also difficult to rule out He-like resonance absorption
in cases where there is an underlying broad Fe K emission line, 
because one does not know how much broad line emission
there is at the resonance energy if some of it has been absorbed.

If the core of the \fekalfa line originates in an accretion
disk we can obtain some simple constraints on the
inclination angle and outer radius of emission given that
the two Doppler peaks from the outer radius of emission are
both contained within the FWHM as measured with
our Gaussian fits. Note that the line profile integrated
between two radii may not actually have discernable
Doppler peaks, but the difference in the energies of the
red and blue Doppler horns at the outer radius still sets
a firm lower limit on the overall line width, assuming azimuthal
symmetry of the line emission. 
Using a simple Schwarzschild geometry
and the approximations in Yaqoob \etal (2003a),
the condition that the line centroid is shifted
by less than $\epsilon \equiv \Delta E/E_{0}$ is
$(r/r_{g}) > 2[1-(1-\epsilon)^2]^{-1}$, where $r_{g} \equiv GM/c^{2}$.
For $\Delta E = 50$~eV and $E_{0}=6.4$~keV, $r>129r_{g}$.
For the worst case, in which the He-like Fe K line is
shifted down to 6.4 keV ($\Delta E=280$~eV, $E_{0}=6.7$~keV),
$r>24r_{g}$. Now, for a given outer radius, if the
disk inclination is too large, the line will be too broad.
The separation of the Doppler peaks from the emission at the
outer radius must be less than the FWHM, so 
$2\sqrt{r_{g}/r} \sin{\theta}  < {\rm FWHM}/c$
(e.g. see Yaqoob \etal 2003a).
Combining this with the energy shift condition gives
$\sin{\theta} < ({\rm FWHM}/c)(2[1-(1-\epsilon)^{2}])^{-\frac{1}{2}}$.
Thus for the group~1 cases in which the lines are unresolved,
using FWHM=$1860 \ \rm \ km \ s^{-1}$ and $\Delta E = 10$~eV
gives a very tight constraint of $\theta <4.5^{\circ}$.
For FWHM=$10,000 \ \rm \ km \ s^{-1}$ 
(more appropriate for 3C~120 and Mkn~509), and $\Delta E$ in the
range 25~eV to 280~eV, the upper limit on $\theta$ is in the range
$15.5^{\circ}$ to $4.6^{\circ}$ respectively.
The constraints on the group~2 sources are obviously much looser.
In general, if there is a significant disk contribution
to the \fekalfa line core, aside from the small inclination
angle, there must be significant line emissivity at large radii.
This would be more easily achieved if the X-ray
continuum source illuminating the disk were extended 
over the disk (for example, the corona) rather than
localized at the center of the system. Whether the
continuum source is centrally localized or extended,
a geometry in which the disk becomes flared at
large radii would also help.

\tablehegfits and \figewvsfwhm show that the EWs of the
\fekalfa line core are typically in the range $\sim 40-200$~eV.
An EW of 40~eV can easily be produced by a column density 
of $10^{23} \ \rm \ cm^{-2}$, covering 35\% 
of the sky as seen from the X-ray continuum source, values that 
are reasonable for the optical BLR in NGC~5548 (see discussion in Yaqoob \etal 2001).
A line core EW of 40~eV could also conceivably be produced by the
outer regions of an accretion disk view at small inclination angles
(e.g. see George \& Fabian 1991). However, values of EW  at the
higher end of the measured range ($\sim 100-200$~eV)
require supersolar iron abundances or anisotropic illumination of
the line-emitting matter by the X-ray continuum. 
Line-emission from a parsec-scale torus structure, that has
been invoked in AGN unfication schemes, could also help
in accounting for the larger EWs since it could subtend
a substantial solid angle at the continuum source and easily
contribute another $\sim 50$~eV to the core EW.
If the torus is optically thick a Compton-reflection continuum
is predicted, commensurate with the EW of the line emission.
In principle this would be a useful observational diagnostic.
However, uncertainties in the Compton-thickness of the
torus, the iron abundance, the contribution to the reflection
continuum from the outer disk, and the amount of line contribution
from the (optically-thin) BLR,
not to mention the measurement uncertainties in the
reflection continuum itself, make it difficult to draw
robust conclusions from correlating the line EW with
the strength of the reflection continuum.
The situation is sufficiently complicated that 
these tests must be done on a source-by-source basis and is beyond the
scope of the present paper.
We note also that the precision of our new measurements of peak energy,
core FWHM, and EW will allow more stringent tests of
alternative
models of the origin of the \fekalfa line emission
than has hitherto
been possible (e.g. see Sulentic \etal 1998; Elvis 2000).

The authors thank Ian George, Jane Turner, 
Barry McKernan, James Reeves, Richard Mushotzky, Sergei Nayakshin,
Kim Weaver, and Peter Serlemitsos
for valuable discussions, and an anonymous referee for helping to improve
the paper.
Support for this work was provided by NASA through \chandra Award Numbers
GO2-2102X, and GO2-3133X issued by the Chandra X-ray Observatory Center,
which is operated by the Smithsonian Astrophysical Observatory for and
on behalf of the NASA under contract NAS8-39073.
The authors also gratefully acknowledge support from
NASA grants NCC5-447 (T.Y., U.P.), and NAG5-10769 (T.Y.).
This research
made use of the HEASARC online data archive services, supported
by NASA/GSFC. This research has made use of the NASA/IPAC Extragalactic Database
(NED) which is operated by the Jet Propulsion Laboratory, California Institute
of Technology, under contract with NASA.
The authors are grateful to the \chandra 
instrument and operations teams for making these observations
possible.

\begin{figure*}[h]
\vspace{-3cm}
\centerline{\psfig{file=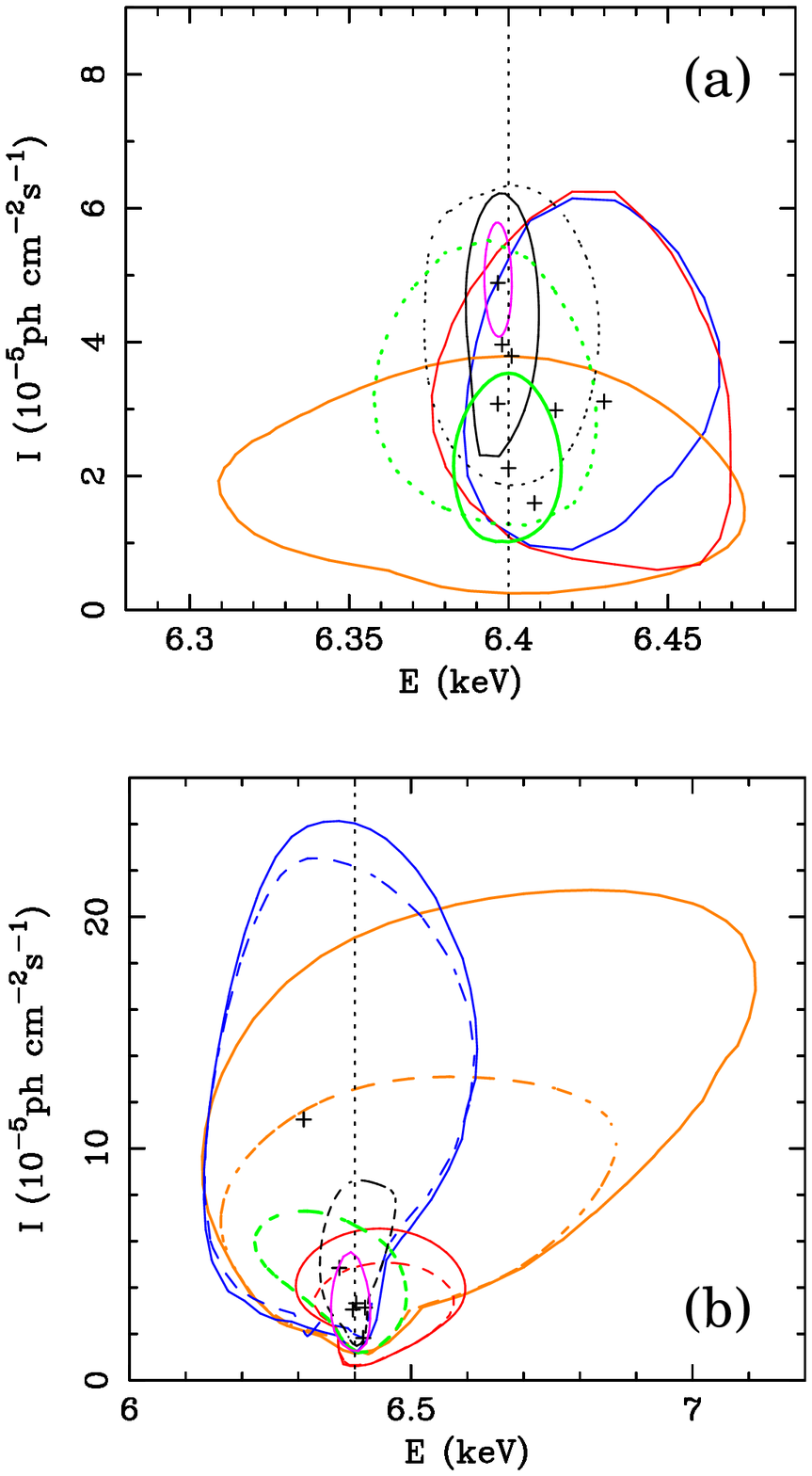,width=6.0in,height=8.0in,angle=180}}
\vspace{-3cm}
\caption{\footnotesize
(a) Joint 99\% confidence contours of the \fekalfa emission-line
core
intensity versus line center energy obtained from 
Gaussian fits to the line as described in the text, for eight
observations of six 
AGN: NGC 3516 (black), Mkn~509 (red), NGC~5548 (green), 3C~120 (blue), 
NGC~3783 (magenta), and
MCG$-$6-30-15 (brown).
See also \tablefits and \figewvsfwhmp.
Dotted contours are for the same source but from a different
observation (the dotted contours are for the lowest
signal-to-noise spectrum of the pair in each case). 
Note that the very smallest contour (magenta) is from the
$\sim 850$~ks observation of NGC~3783.
The first observation of NGC~3783 was excluded from the plot
because it is has a factor of $\sim 20$ less exposure time.
Also excluded were NGC~7314 (Fe K line
emission was very complex), and Mkn~766, NGC~3227, and Akn~564, 
that all had insufficient signal-to-noise to obtain well-constrained
contours.
(b) As (a), for five more AGN: NGC~4593 (black), Mkn~279 (red), 
NGC~4051 (green), IC~4329a (blue), 
and F~9 (brown).
Also shown here is the 99\% contour for NGC~5548(1) (magenta, 
and the smallest contour in
the plot)
to compare with the contour of F~9. These two data sets have
a similar number of counts in the 5--7~keV spectra
but the size and shape of the contours are completely different.
This shows that the 
differences in the size and shape of the contours
in general reflect intrinsic differences in the line profile shapes.
The dashed contours were obtained by fixing the power-law
continuum slope after finding the best fit because
closed contours could not be obtained otherwise for NGC~4593 (black) and 
NGC~4051 (green).
For the sources in \figivse(a), the contours obtained by freezing the
continuum slopes at the best-fitting values were not
significantly different from
the fits with all parameteres free, so they are not shown for clarity.
}
\end{figure*}

\begin{figure*}[h]
\vspace{-3cm}
\centerline{\psfig{file=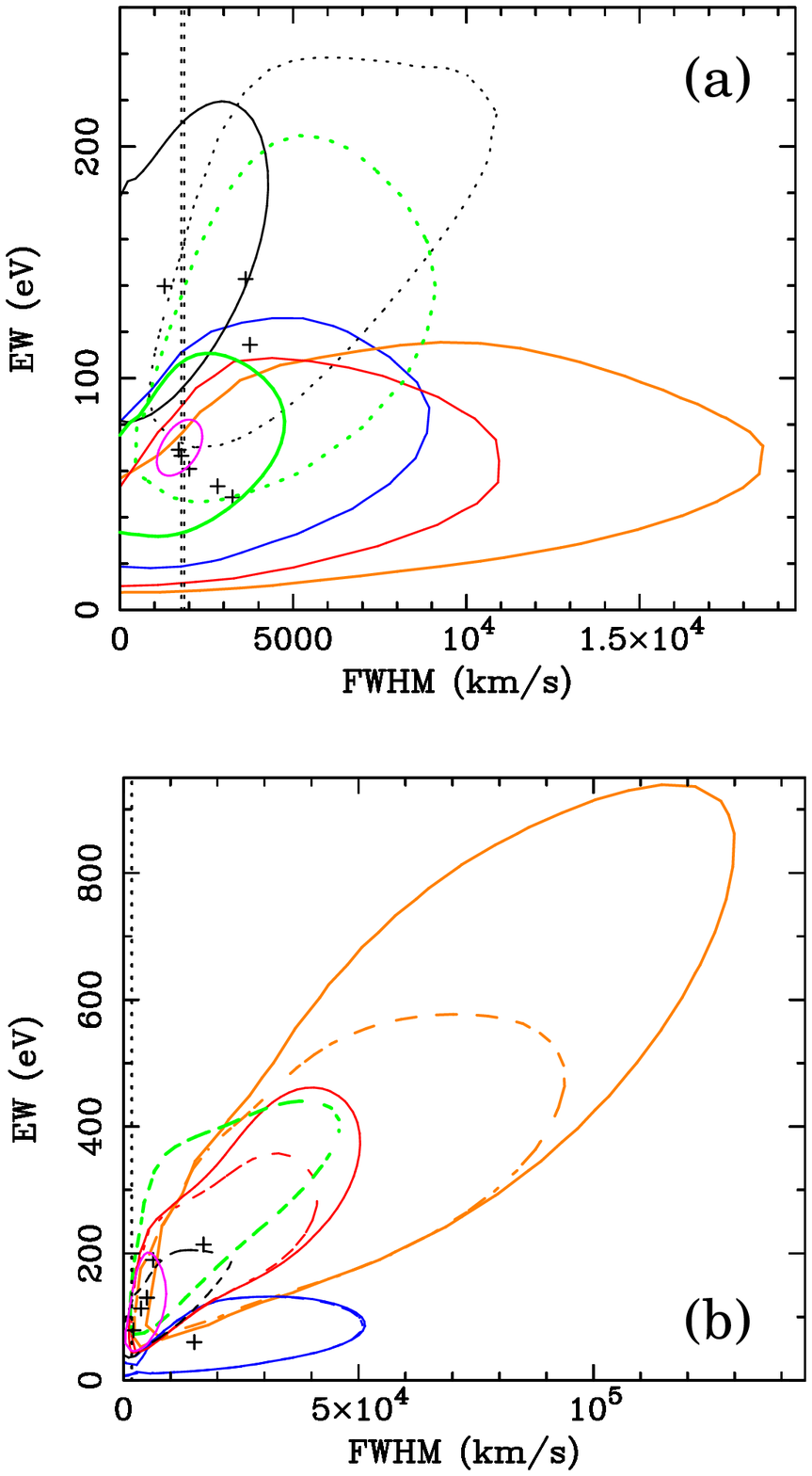,width=6.0in,height=8.0in,angle=180}}
\vspace{-3cm}
\caption{\footnotesize
(a) Joint 99\% confidence contours of the \fekalfa emission-line
core equivalent width (EW) versus velocity width FWHM, for the
same six AGN as in \figivse(a), using the same color coding.
(b) As (a) but for the same five AGN as in \figivse(b), using the same color coding.
Note that the FWHM was calculated simply from $v/c = 2.35\sigma/E_{0}$,
where $E_{0}$ is the line center energy and $\sigma$ is the
Gaussian line width. Strictly speaking, this is valid
only for $v/c \ll 1$ so the conversion is not accurate for 
the highest velocity parts of the contours.
Again NGC~5548(1) from \figewvsfwhm(a) is included for comparison
(small, magenta contour). The dashed contours were obtained
by freezing the power-law continuum slope at the best-fitting value:
for NGC~4593 (black) and NGC~4051 (green) closed contours could
only be obtained in this way.
See also caption to \figivse(b).
}
\end{figure*}

\begin{figure*}[tbh]
\vspace{10pt}
\centerline{\psfig{file=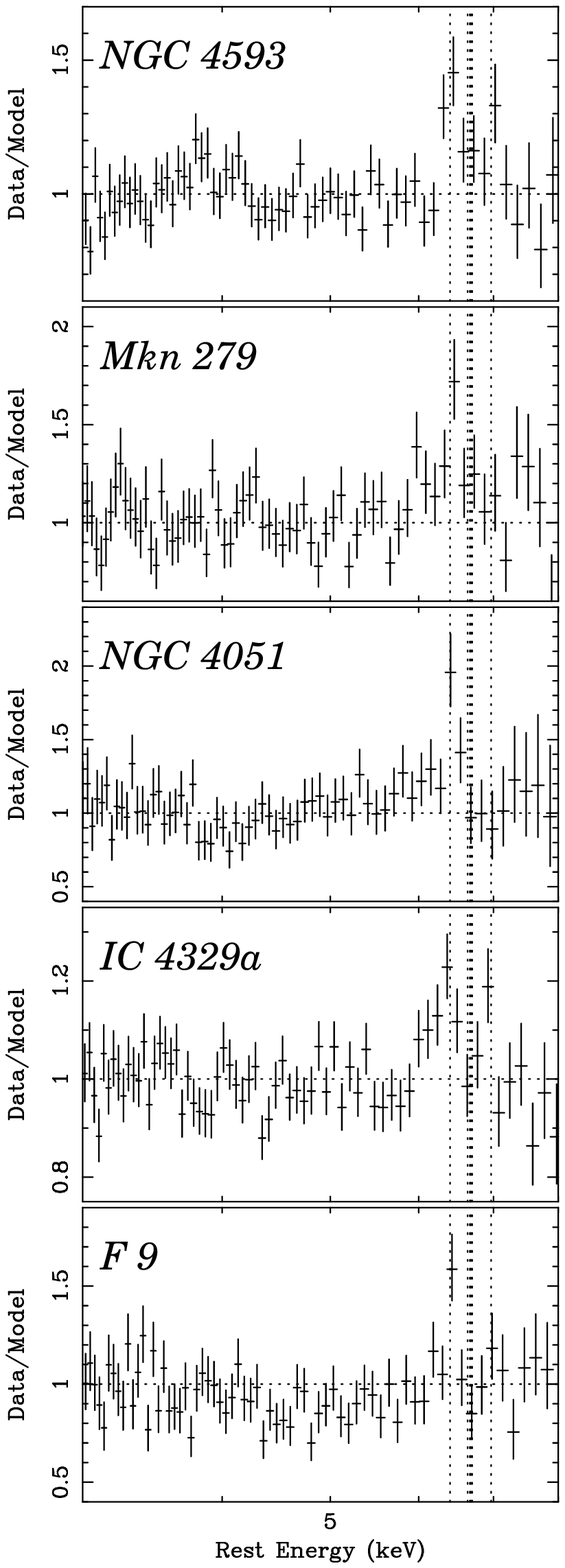,width=3.5in,height=7.5in,angle=180}}
\caption{\footnotesize
Ratios of HEG data to a simple power-law continuum model (fitted as described in the
text, \S\ref{hegspec}) for five AGN in \figfinespec and \tablefits which show evidence of
broad and/or complex Fe K line emission. MCG~$-$6-30-15 is not shown, since
a similar plot has already been shown in Lee \etal (2002). 
The data are binned at $0.04\AA$, coarser than the
HEG spectral resolution ($0.012\AA$
FWHM). The data are combined from the $-1$ and $+1$ orders
of the grating. The spectra have been corrected for instrumental effective area
and cosmological redshift.
The statistical errors shown correspond to the $1\sigma$ 
Poisson errors, calculated using 
equations (7) and (14) in Geherls (1986) to approximate the upper and
lower errors respectively.
The vertical dotted lines represent (from left to right), the rest energies
of the following transitions: Fe~{\sc i}~$K\alpha$, \fexxv forbidden,
two intercombination lines of \fexxvp, \fexxv resonance, and \feklyap.
}
\end{figure*}

\end{document}